# WIRE: Write Energy Reduction via Encoding in Phase Change Main Memories (PCM)

Mahek Desai[1], Apoorva Rumale[1], Dr. Marjan Asadinia[1], Sherrene Bogle[2]

[1] California State University, Northridge CA 91330, USA
[2] California State Polytechnic University, Humboldt
mahek-trushit.desai.849@my.csun.edu
apoorva-sanjay.rumale.462@my.csun.edu
marjan.asadinia@csun.edu
sherrene.bogle@humboldt.edu

**Abstract.** Phase Change Memory (PCM) has rapidly progressed and surpassed Dynamic Random-Access Memory (DRAM) in terms of scalability and standby energy efficiency. Altering a PCM cell's state during writes demands substantial energy, posing a significant challenge to PCM's role as the primary main memory. Prior research has explored methods to reduce write energy consumption, including the elimination of redundant writes, minimizing cell writes, and employing compact row buffers for filtering PCM main memory accesses. However, these techniques had certain drawbacks like bit-wise comparison of the stored values, preemptive updates increasing write cycles, and poor endurance. In this paper, we propose WIRE, a new coding mechanism through which most write operations force a maximum of one-bit flip. In this coding-based data storage method, we look at the frequent value stack and assign a code word to the most frequent values such that they have a hamming distance of one. In most of the write accesses, writing a value needs one or fewer bit flips which can save considerable write energy. This technique can be augmented with a wear-leveling mechanism at the block level, and rotating the difference bit in the assigned codes, increasing the lifetime of the PCM array at a low cost. Using a full-system evaluation of our method and comparing it to the existing mechanisms, our experimental results for multi-threaded and multi-programmed workloads revealed considerable improvement in lifetime and write energy as well as bit flip reduction.

## 1 Introduction

Among all Non-Volatile Memories (NVMs), PCM is considered the leading contender for next generation of main memory system [1-4]. PCM cell exploits Chalcogenide material (GST) and shows low resistive crystalline state (set state) and high resistive amorphous state (reset state). PCM has appealing characteristics such as high scalability, non-volatility, low leakage power, and reasonable read latency. Despite its benefits, PCM suffers from limited write endurance and can only tolerate a limited number of write operations (which is about $10^7$-$10^9$ writes per cell before wearing out).

Upon reaching the lifetime limit, it becomes stuck-at "1" or "0" and results in low cell reliability [2,4]. To overcome PCM lifetime limitation, existing solutions are grouped in two categories: 1) bit flip reduction schemes which can be orthogonally used with wear-leveling techniques. These techniques try to reduce the number of bit flips per write request and can reduce write rate to PCM cells by uniformly spreading write traffics among cells. 2) handling cell failures when faults occur (using error detection and error correction codes) and postponing the wearing out of PCM cells [1-3,14, 15]. For the first category, the simplest bit flip reduction scheme is presented in [7] and called a differential write. This method relies on a bit-by-bit comparison between the currently stored data in the memory and new data to be written. In the case that stored bit values differ from the new bit values, writing to the cell is invoked. Otherwise, differential write skips unnecessary bit programming. Clearly, if there is more similarity between the new data and the currently stored data, less bits are needed to change and power consumption can be saved. So, differential write faces a problem in the case of either low similarity or no similarity. Flip-N-Write is proposed [6] to further reduce bit flips and solve the lower similarity problem in differential write scheme. This method exploits one additional bit, called a flip bit per each word, to identify which part of the data is inverted and which one is not inverted. It then calculates the hamming distance of the stored data and new data to be written. In this method, N is the data bit width and the input data is inverted if the calculated hamming distance is more than N/2. This approach selectively inverts the blocks of data to be written, and thus provides more similarity. The major overhead of Flip-N-Write is extra read for issuing the exact data as well as the additional bit flip associated with each PCM word. To this effect, a cost model, CAFO [8], is presented to compute the cost of servicing write requests through assigning different costs to each cell that requires programming. CAFO encodes the data to be written into a form that imposes less cost through its encoding module.

To address the shortcoming of existing methods, we demonstrate that the PCM write mechanism as well as the bit flip problem bring negative impacts on write access energy when reliable and dense PCM memories are required. We propose a novel coding mechanism for bit flip reduction and lifetime extension of PCM with data encoding, named WIRE, Write Energy Reduction via Encoding in Phase Change Main Memories (PCM). Based on our observation, there is a nonuniform data pattern distribution of memory transactions known as frequent value locality. Therefore, we develop a frequent value finder logic that can identify relative occurrence frequency of data patterns and detect most frequent value locality stack (MFVs). Based on frequent value locality phenomenon, a small set of distinct values span a large fraction of fetched/stored data within a memory system and there is a high degree of replication for accessed values within a storage system. We then design a new coding mechanism that assigns specific codes to the identified MFVs with the hamming distance of one. The proposed code guarantees fast access and reduces the bit flip problem. On the other hand, a high-level interface for value translation is presented to facilitate value translation in operating system. We are optimizing our coding mechanism by partitioning and rotating the assigned codes with the goal of finding more similarity between the data patterns and get more improvement in bit flip reduction.

## 2   Related Works

In the expansive domain of computer architecture research, a diverse range of energy-saving methodologies has been developed, with particular emphasis on Phase-Change Memory (PCM) systems. These methodologies encompass a broad spectrum of innovations, spanning from foundational advancements in phase-changing states to intricate approaches in data management and system optimization [2], [5]. Notably, the application of machine learning and predictive analytics has emerged as a promising avenue to forecast and prefetch frequently accessed values, thereby effectively minimizing unnecessary energy expenditure [7]. However, it's crucial to recognize certain limitations within these methodologies. While machine learning and predictive analytics offer the potential to forecast and prefetch data, their accuracy heavily relies on the quality of training data and the complexity of the workload patterns. Moreover, the computational overhead associated with training and inference processes may offset the energy savings achieved through prefetching [11], [13].

Beyond these advancements, significant attention has been directed towards refining the operational intricacies of PCM systems, particularly in minimizing write cycles and optimizing energy consumption during write operations. Strategies such as buffering small writes and combining adjacent writes have been extensively investigated, showcasing potential to significantly reduce write energy consumption [4], [6]. However, it's important to note that these strategies may introduce additional latency and complexity to the write process, potentially impacting overall system performance.

Additionally, innovative techniques like the Partial Write and Multi-Read (PWMR) method have further augmented efforts towards enhancing energy efficiency and overall system throughput [4], [12]. Despite their promise, these techniques may face challenges in implementation complexity and compatibility with existing memory architectures, potentially limiting their widespread adoption and effectiveness in heterogeneous computing environments.

Moreover, other research endeavors have proposed innovative schemes such as the three-stage-write scheme with flip-bit technique [10], dynamic modification of write modes based on data characteristics through compiler-guided optimization [11], and the implementation of write operation batching techniques to exploit parallelism effectively within PCM-SSD architecture [11-15]. However, these schemes may encounter challenges in terms of implementation overhead, compatibility with existing software stacks, and scalability across diverse computing platforms. In addition to these efforts, a concerted effort is observed in optimizing write operations through techniques such as data grouping and write coalescing for efficient storage, all aimed at further mitigating write energy consumption [16, 17]. While these techniques offer potential energy savings, they may introduce complexity in memory management and data organization, potentially increasing overhead in terms of memory access latency and system resource utilization. Additionally, their effectiveness may vary depending on workload characteristics and system configurations, making their adoption and deployment non-trivial in real-world computing environments [10-16]. Collectively, while these multifaceted research endeavors represent significant progress towards minimizing write energy and enhancing overall performance in PCM main memory systems, it's essential to acknowledge the inherent challenges and limitations associated with each method. Addressing these weaknesses will be critical in advancing the field towards more efficient and scalable energy-saving solutions in computer architecture research.

## 3  Proposed Method

In this section, we first present our frequent value finder logic to detect the most frequent values within a memory system. We then explain our coding mechanism.

### 3.1  Frequent Value Finder Logic

We first study the behavior of multi-threaded PARSEC-2 benchmarks [20] to establish the existence of frequent value locality in real programs. We use main memory traces of PARSEC-2 and capture them from a full-system simulation (we show the system specification in Table I). After program execution, we identify the frequently encountered values by tracking the execution of all fetch and store instructions. In this experiment, we examine values that are frequently accessed by memory operations and show the frequency of those values from most to least frequent. To better understand, we use 4-bit granularity and show the percentage of 16 different values read/written from/into main memory.

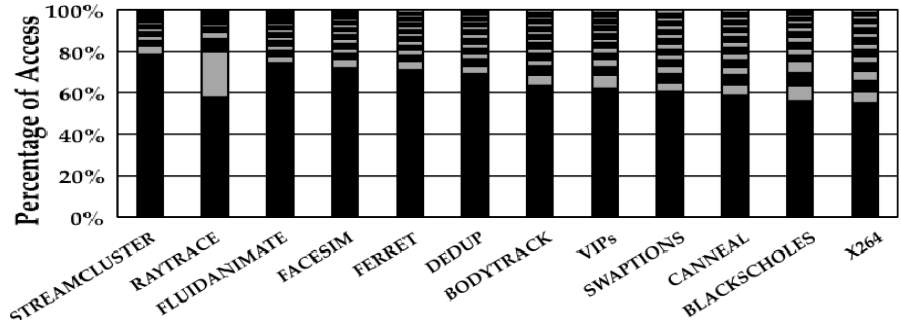

**Fig. 1.** Frequency of accessed values from most to least accessed for 16 different values.

As it is shown in Fig. 1, five leftmost programs exhibit about 80% contribution of most frequent values, next four studied programs contain 65% - 70% of such values, while the last three programs contain about 65% frequent values. Since we want to extract frequent values during initial phases of the program, a monitoring scheme is the necessity of our approach.

Based on past observations, we found that single threaded workloads follow stable and similar frequency at different run time intervals. On the other hand, multiprogram workloads consist of two types of patterns. One type has the common patterns, like all-zero and all-one patterns. The second type has different values among various applications and, most frequent values change from application to application. Indeed, multiprogram workloads follow dynamic and variable frequency during the occurrence of different values. Fig. 2 shows the data pattern distribution in benchmark programs.

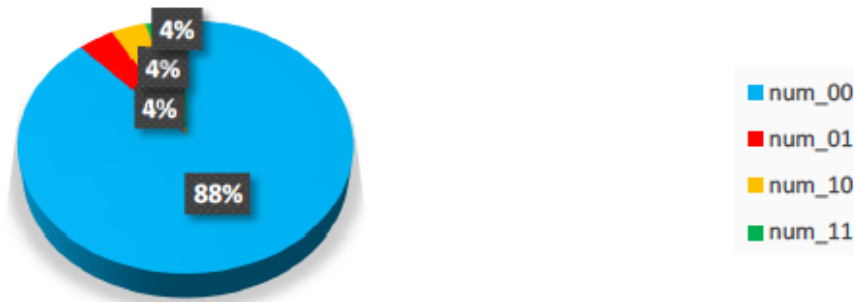

**Fig. 2.** Data Pattern Distribution in benchmark programs.

To this end, we need a fast and low-overhead PCM access circuitry to support the extraction of most frequent value locality stack without negative effects on the memory access critical path.

The Most Frequent Value Logic (MFV) consists of two separate entities. The first is the FIFO buffer, which filters out transient occurrences of non-frequent values and retains only the frequent ones. The buffer has three fields: a value field, a data field and a saturation counter field to track and distinguish data. When a data pattern is encountered this FIFO buffer is searched and if the data is found (hit), the corresponding counter is increased and can be considered a frequent value if saturation is met. Conversely, if the data is not found (miss), all counters are decremented by one and if a counter is less than a threshold value, its entry is replaced by a new pattern.

To store the value and code word of the frequent patterns another table called Frequent Value (FV) table is used. The table has four fields: a value field, a counter field, a pointer field and a used bit field. The counter is inherently an incrementing one that increases on each frequent value access. A line in the table is marked as a Gap that does not refer to any frequent value i.e. used bit=0. If a value in the FIFO buffer is marked as a frequent pattern and subsequently there is a Gap in the frequent value table then the value from the FIFO buffer is replaced in the Gap line with the used bit set to 1. As the memory blocks are transferred in and out, the pointer field of the Gap line decreases, and its used bit is reset when the pointer field becomes zero. Note that for efficient operation of the MFV logic, the counter field of the frequent value table must be longer than the FIFO buffer. Synthesis shows frequent value finder logic imposes a negligible power overhead of 22mW (about 1%) to the overall power consumption of the PCM main memory array.

**Table 1.** FIFO Table.

|  | Entry | Value | Saturation counter |  |
|---|---|---|---|---|
| hit | Entry 0 |  | 1 |  |
|  | . |  |  |  |
|  | . |  |  |  |
|  | . |  |  |  |
| miss | Entry n |  | 0 | < threshold |

**Table 2.** FV Table.

|  | Entry | Value | Counter field | Pointer field | Used bit |
|---|---|---|---|---|---|
| Gap line | Entry 0 |  |  |  | 0 |
|  | . |  |  |  |  |
|  | . |  |  |  |  |
|  | . |  |  |  |  |
| Victim; if CTR < threshold | Entry n |  |  |  |  |

All of the above explanations are summarized in Table 1 shows the FIFO table and its different fields. Table 2 shows the operation and different fields of the frequent value table.

### 3.2 Assigning specific codeword to Most Frequent Values

Upon detection of Frequent Values (MFVs), and subsequent assignment of specific codewords to them based on their data patterns, our coding-based data storage scheme, known as Write Energy Reduction via Encoding (WIRE), facilitates access to other upcoming and new data patterns through the assigned codewords to the MFVs of the stored data. To minimize bit flips, WIRE only accepts a Hamming distance of one between two codewords for different data patterns.

For example, we consistently designate the codeword "0" to represent all zero data patterns and the codeword "1" for all one data patterns. When confronted with a combination of zero and one data patterns, our coding mechanism explores various permutations and assigns the codeword based on the previously stored MFV. Should the codeword of stored MFVs fail to grant access to other data patterns with a Hamming distance of one for writing in the memory, our memory controller initiates partitioning and rotation of the codewords.

In this process, the memory controller first evaluates the Hamming distance between the new data and the stored data to ascertain any similarity. Based on this distance, it determines the number of rotations required for codeword adjustment. We adopt fixed and static partitioning, typically in partitions of 8 for a 64-byte data block, to circumvent the complexities and power consumption associated with dynamic partitioning. Additionally, we cap the number of rotations at 8 bits, a value determined through sensitivity analysis to ensure reasonable storage overhead. More precisely, we can use this rotation scheme in each partition for the currently stored value such as MFV1 pattern; and can achieve the MFV2 data pattern or other MFVs by up to 8 bits rotation.

To better understand, assume that we have only a 4-bit data pattern, in the first memory transaction the data pattern with the value of "1001" comes up, our frequent value finder logic detects it as one of the frequent data patterns which has a combination of zero and ones. So, it considers MFV1 for this data pattern and assigns the specific code word to MFV1. Since we have only a 4-bit data pattern in this example, assume our assigned codeword is also "1001". After a while, some memory blocks are transferred in and out of memory, some of them can be selected as frequent values and frequent value finder logic assigns MFVs to the corresponding data patterns. At this time, assume another data pattern is transferred to the memory controller with the data pattern of '1000' and requests a write process. Our memory controller assigns the new codeword to this data, compares its codeword to the stored codeword of the frequent data pattern (other MFVs), and then detects that it has the hamming distance of one with MFV1 because they have only a one-bit difference. Therefore, we can easily write this data pattern in the memory blocks with low write energy and low number of bit flips by using MFV1. In the next memory transaction, if we get another data with the pattern of '0010', our memory controller again searches to find some MFVs that match this upcoming data pattern, or some MFVs with a hamming distance of one to this current data pattern. If it fails and the memory controller cannot find a matching option, it invokes our rotation mechanism. Based on checking the hamming distance of the stored data pattern and the new data pattern to be written, a 2-bit difference is determined. Therefore, our coding scheme rotates the stored data by up to 2 bits to reach the new data pattern. If we have more than 4-bit data patterns, like a 64B data block, we should partition the data block and then exploit the rotation in each partition, same as we do in the above assumption.

By augmenting the rotation scheme to our coding mechanism, we can get more savings in write energy due to the reduction of bit flips. All of the assumptions and the validity of this proposed technique will be demonstrated through our sensitivity analysis. This low overhead scheme prevents non-uniformity in bit flips across memory blocks. Also, we can remove the stress from hot locations and relax the worst-case bit flip rates in each block.

During read operations, if a read request is issued to the memory controller, it initially determines whether the data block corresponds to the MFV. If affirmative, a standard read iteration suffices to deliver the data; otherwise, the read circuit retrieves the content of the previous block and, based on its codeword, obtains the correct content of the block.

For write operations, upon receiving a write request, the scheme checks the frequent value finder logic to ascertain whether it is the MFV. It then determines the codeword of the block and rotates it based on the Hamming distance to find a similar or nearly identical codeword. Once a suitable rotated codeword is identified, the original codeword is retrieved, and the data is sent to the write circuit to complete the operation.

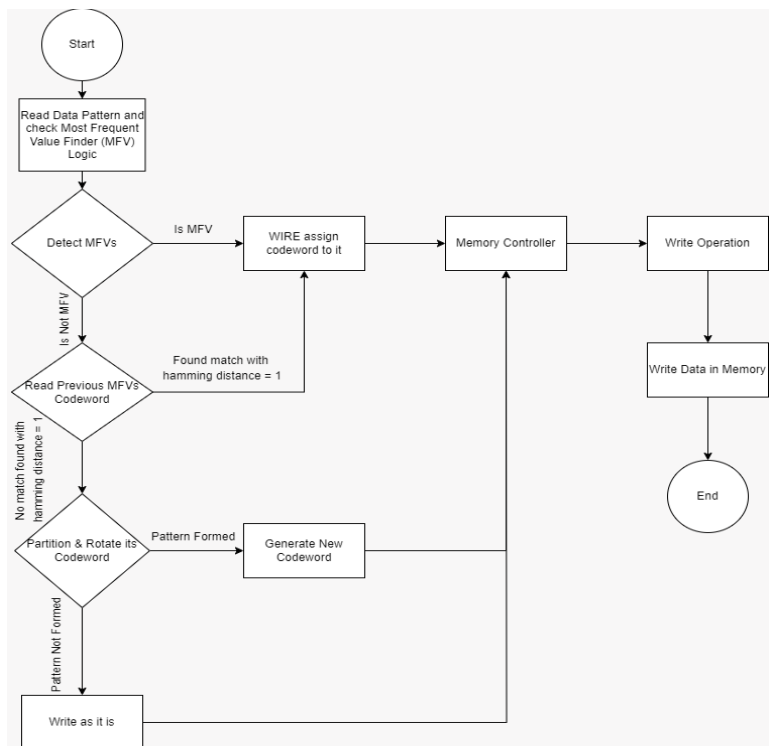

**Fig. 3.** WIRE Logic Block

Our WIRE scheme employs the rotation mechanism when encountering new data patterns without similar codewords with a Hamming distance of one. By partitioning the stored codeword and rotating it based on the Hamming distance between the new data and the stored data, we can identify similarities with a tolerable amount of rotation. The logical blocks of WIRE are illustrated in Fig. 3. It is notable that we use the rotation scheme when we have the new data pattern and cannot find any similar codewords with a hamming distance of one. Then, we first partition the stored codeword and rotate it based on the hamming distance of new data and stored data. By a tolerable amount of rotation, we can find some similarities between new data to be written and the stored data.

### 3.3 OS Support

We require operating system support for our proposed approach. Specifically, a workload-based value translation mechanism is essential to accommodate variations in the most frequent values. Let's consider a scenario where the first most frequent value (MFV1) for workload1 is '100' and for workload2 it's '101'. If both are assigned to the same PCM cell with identical resistance levels, we need a translation mechanism. This process can be managed by the operating system when multiple applications are running. The operating system maintains awareness of the owner of each physical memory page in the memory management system. This ownership mechanism ensures privacy at the OS level.

## 4 Evaluation

In this section, we provide a brief overview of simulation parameters, workloads, and selected performance metrics.

### 4.1 Evaluation Environment

| | |
|---|---|
| Processor | 4-core SPARCIII, 4.0GHz |
| L1 Cache | Split I and D cache; 32KB private; 4-way; 64B line size; LRU; write-back; 1 port; 2ns latency. |
| L1 Coherency | MOESI directory; 4×2 grid packet switched NoC; XY routing; 3 cycle router; 1 cycle link. |
| L2 Cache | 4MB; UCA shared; 16-way; 64B line size; LRU; write-back; 8 ports; 4ns latency. |
| DRAM Cache | 16MB; 4-way; 64B line size; LRU; write-back; 8 ports; 26ns latency. |
| Main Memory | 8 GB: 16 banks, 64 B, open page, SLC: Read Latency 80 ns (6ns tPRE + 69ns tSENSE + 5ns tBUS), Write Latency 250ns. |
| Flash SSD | 25μs latency. |

We implemented our proposed method in GEM5+NVMain hybrid full system simulator [18]. A 4-core 2GHz CMP system with 3 levels of caches and 8GB PCM main memory with 16 banks are modeled in our simulator. Our simulation parameters are shown in Table 3.

**Table 4.** Characteristics of the PARSEC-2 workloads.

| Workload | WPKI | RPKI | Workload | WPKI | RPKI |
|---|---|---|---|---|---|
| Blackscholes | 0.003 | 0.03 | Bodytrack | 0.003 | 0.03 |
| Ceneal | 1.12 | 1.13 | Dedup | 0.41 | 0.43 |
| Facesim | 0.65 | 0.24 | Ferret | 0.65 | 0.67 |
| Fluidanimate | 0.76 | 0.47 | Freqmine | 0.04 | 0.07 |
| Raytrace | 0.008 | 0.02 | Streamcluster | 0.01 | 0.07 |
| Swaptions | 0.002 | 0.02 | Vips | 0.05 | 0.07 |
| X264 | 0.04 | 0.04 | | | |

**Table 5.** Characteristics of the SPEC-CPU workload.

| Workload | WPKI | RPKI | Workload | WPKI | RPKI |
|---|---|---|---|---|---|
| Mix 1 | 2.4 | 2.23 | Mix 6 | 6.64 | 10.11 |
| Mix 2 | 3.69 | 3.08 | Mix 7 | 1.99 | 0.64 |
| Mix 3 | 2.2 | 0.8 | Mix 8 | 1.91 | 2.48 |
| Mix 4 | 0.27 | 0.25 | Mix 9 | 0.53 | 1.08 |
| Mix 5 | 0.53 | 0.66 | | | |

**Real Workload.** We use multi-programmed workloads from SPECCPU-2006 [19] and a complete set of parallel workloads provided in PARSEC-2 [20] as multi-threaded workloads to evaluate different methods.

**Evaluated Architectures**. We compare our proposed method against two bit flip reduction schemes, Differential Write method [9] and Flip-N-Write mechanism [8]. Our evaluation metrics include memory capacity degradation, write energy and IntraV [21].

### 4.2 Analysis Under Real Application

**Capacity degradation (Lifetime)**. We define the lifetime as time to failure metric. We measure the elapsed time between the start time and the time when half of memory pages worn out and the memory capacity degrades to 50%. So, we simulate each application to the defined time and measure the elapsed time for the evaluated systems. Fig. 4 illustrates our method provides better memory capacity since it improves the bit flip reduction and postpones the wear out of PCM block by using coding mechanism.

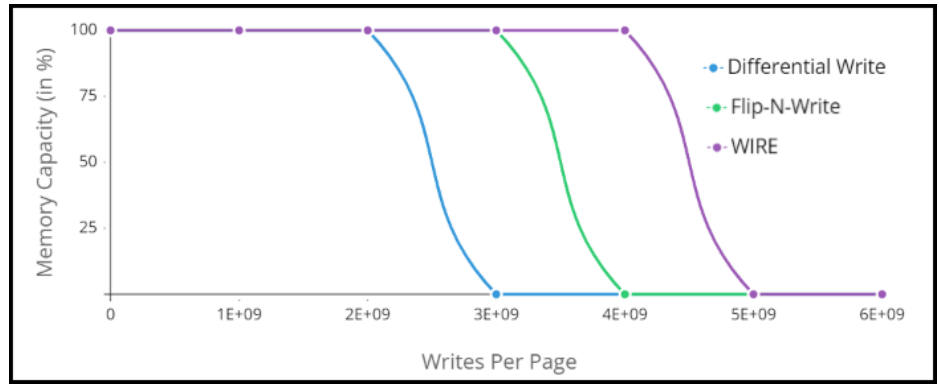

**Fig. 4.** Lifetime of WIRE compared to other techniques.

**Bit flip reduction**. For evaluating the bit flip reduction, we use the parameter, called IntraV which is defined in [21].

$$IntraV = \frac{1}{BF_{aver}.N} \times \sum_{i=1}^{N} \sqrt{\frac{\sum_{j=1}^{512}(BF_{ij} - \frac{\sum_{j=1}^{512} W_{ij}}{512})^2}{511}} \quad (1)$$

where the write count of cell *j* in block *i* is shown as $BF_{ij}$, the average bit flips count is $BF_{aver}$ and the total number of blocks is *N*. Where $BF_{ij}$ is the write count of cell j in block i, and $BF_{aver}$ is the average bit flips count and N is total number of blocks.

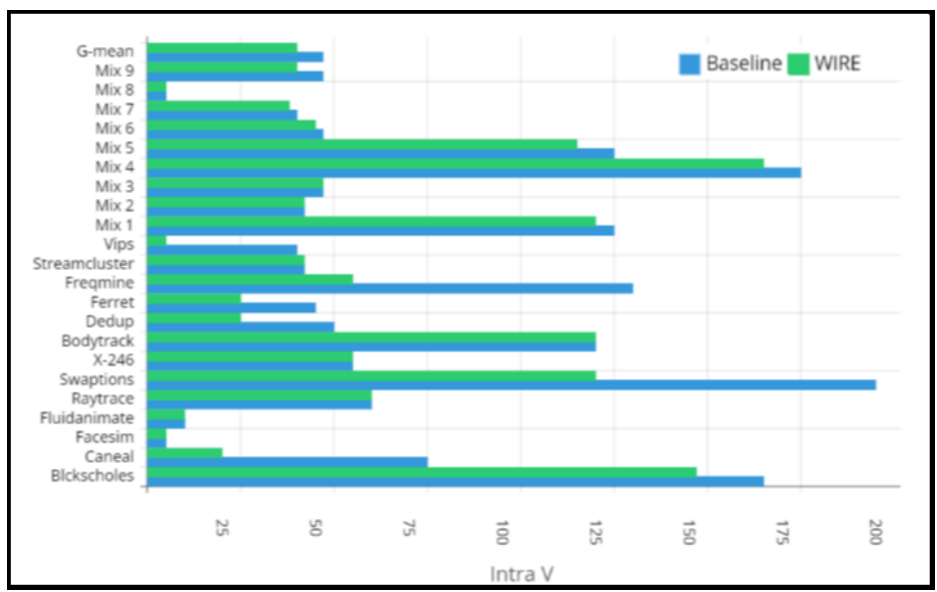

**Fig. 5.** IntraV comparison of WIRE and baseline.

Fig.5 shows the IntraV of the workloads before and after using WIRE. We consider the amount of reduction in bit flips for different multi-thread and multi-program workloads by using this parameter. Indeed, our scheme can reduce the variation of bit flips in each block by 25%, on average.

**Write Energy**. Fig.6 shows energy consumption improvement of our proposed method compared to the other techniques. Since we use frequent value locality phenomena in our coding-based scheme, we have better write energy due to bit flip reduction and presence of more similarity in the data patterns. However, differential write imposes high write energy because it uses a bit-by-bit comparison technique. Conversely, Flip-N-write has a better write energy against differential write technique (inversion technique of Flip-N-Write provides this improvement compared to the differential write). WIRE results in an improvement of 24% - 32% in write energy.

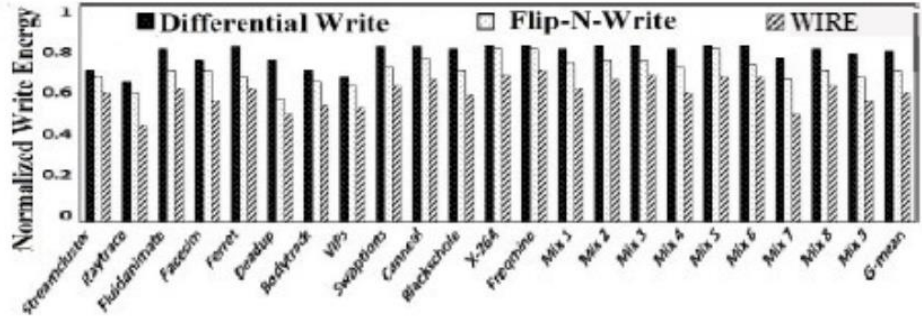

**Fig. 6.** Total write energy of WIRE and other techniques normalized to the conventional PCM.

**Sensitivity analysis on partition size and meta-data information**. When the assigned codeword to the stored data has the hamming distance greater than one, we optimize our coding scheme and use the partition and rotation for the codeword to achieve more similarity and closer hamming distance. We believe that the best partitioning size in WIRE plays an important role in determining the storage overhead and IntraV value. So due to limitation for write energy consumption, we use the fixed partitioning instead of the dynamic partitioning. On the other hand, larger partitions have a smaller number of units which reduce IntraV parameter, and it is not desirable. Based on the impact of partition size on IntraV. We select 8 partitions for each data block of 64B which provides a good trade-off for IntraV. In fact, for rotation part, we may need 6-bit counter for every partition. This configuration facilitates our shift-based mechanism, aimed at finding more similarity with the reasonable hamming distance. Additionally, these counters serve to indicate the number of shifts applied to block content when encountering no match with assigned codeword. Consequently, we require $6 \times 8 = 48$ additional bits per block, resulting in a storage overhead of $48/512 = 9\%$.

In this case, the shifting mechanism assumes a pivotal role in overhead. However, to ensure practicality during read or write operations, we have refrained from enlarging the row size of the memory to accommodate counter values for each line. Instead, we opt to store counter values in alternative memory lines, necessitating an extra read operation to retrieve these values. Alternatively, one could integrate counter values into a memory line, thereby extending its length through additional memory banks, allowing for the simultaneous read or write of an entire line (data block and meta-data counter values) within one memory cycle. Nevertheless, storing counter values in separate memory lines introduces latency overhead due to the additional read operation required to obtain these values.

To mitigate the adverse effects of this arrangement, we can employ a caching mechanism akin to the Failed-Cache concept in previous proposals. This mechanism aims to reduce the latency overhead resulting from the extra read operation required to access counter values. Essentially, a fraction of our meta-data is stored in a small cache within the memory controller. This ensures that for the majority of memory accesses, counter values are readily available in the cache, thereby minimizing latency overhead. For instance, with a 48-bit overhead per block, a 2KB cache can accommodate the meta-data of (2048/6) = 341 blocks, which proves satisfactory.

### 4.3  Analysis Under Synthetic Application

We conducted experiments to measure the total energy consumption of PCM under various read and write conditions by using synthetic trace files generated by NVMain. We considered three cases in our evaluation: 1) when the number of read operations is greater than write operations, 2) when we have equal demands for both read and write operations, and 3) when there is a heavy bias towards writes over reads.

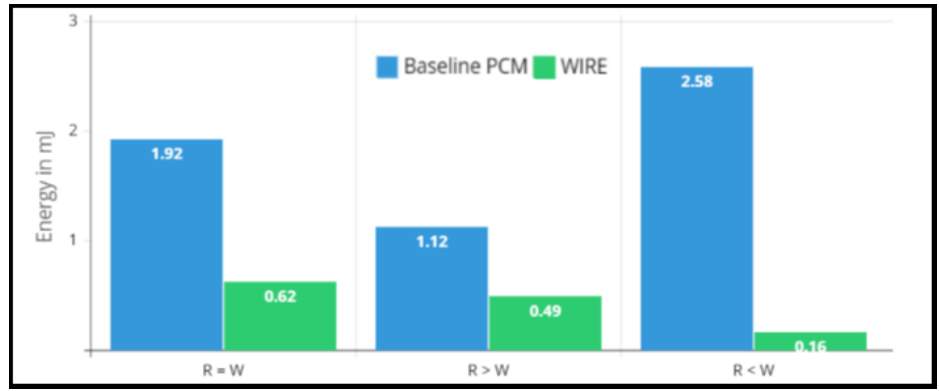

**Fig. 7.** Write energy comparison under various read and write conditions.

**Write Energy Analysis**. We evaluated the energy dedicated solely to write operations. Through detailed analysis, we observed significant reductions in write energy consumption with WIRE compared to the baseline PCM, as seen in Fig.7. This reduction in write energy is attributed to the innovative encoding scheme proposed in WIRE PCM, which optimizes write operations to minimize energy consumption while ensuring data integrity.

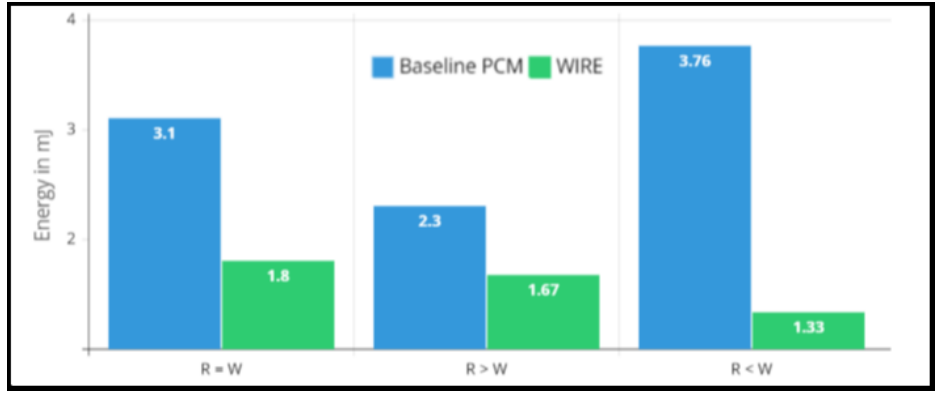

**Fig. 8.** Total energy comparison under various read and write conditions.

**Total Energy Analysis**. Fig.8 indicates a substantial reduction in total energy consumption when employing WIRE PCM compared to the baseline PCM. Specifically, we observed reductions in total energy across all scenarios, demonstrating the effectiveness of our proposed method in improving energy efficiency in synthetic application environments.

## 5   Hardware Overhead and Extension for N-bit MLC PCM

In Figure 10, the schematic illustrates the access logic supporting Single-Level Cell (SLC) and 2-bit Multi-Level Cell (MLC) read/write operations. Primarily, two components significantly contribute to the area of the access circuit: (1) sense amplifiers and (2) NMOS transistors within the write circuit. Notably, NMOS transistors hold greater significance in Phase Change Memory (PCM) systems due to their role in driving substantial currents into cell arrays. Area calculations conducted for the access circuit in a 45nm technology, utilizing HSPICE and CACTI, reveal that incorporating additional logic to accommodate MLC operations incurs a 30.4% overhead compared to SLC's access logic. However, this overhead represents less than 0.5% of the PCM chip's total area. Given the escalating demand for larger main memory capacities across various computing domains, SLC PCM memory might become inadequate, consequently fostering a preference for PCM memory systems with higher bit densities. Numerous prior studies have advocated for 2-bit MLC PCM as the preferred technology for main memory systems.

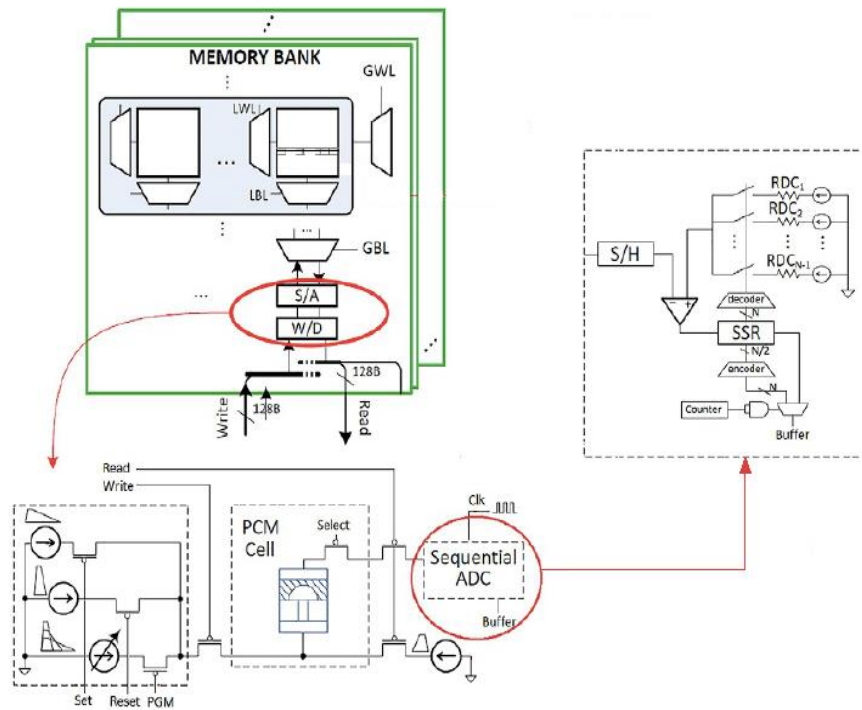

**Fig. 9.** The PCM cell that supports N-bit to 2N-bit MLC operations.

Contrarily, the WIRE model posits a memory system commencing with SLC storage but remains adaptable to transitioning to a 2-bit MLC storage level, thereby ensuring scalability. Consequently, read/write circuits must support both SLC and 2-bit MLC accesses, resulting in a minor circuit overhead. As depicted in Figure 9, executing an MLC PCM write necessitates a full-sweep RESET pulse (for programming into RESET state or initializing the GST for a Program and Verify process), a full-sweep SET pulse (for programming into SET state), and a Programmable pulse (PGM) facilitating partial SET pulses during Program and Verify operations. The PGM unit integrates a current Digital-Analog Converter (DAC) capable of generating SET pulses with specified widths but varying amplitudes. To elevate the density of an N-bit MLC PCM to that of a 2N-bit MLC PCM, the PGM control logic determines the amplitude of intermediate SET pulses. Conversely, for reads, the controller adjusts the sequential Analog-to-Digital Converter (ADC) iterations according to the targeted storage level. Therefore, transitioning from an N-bit MLC to 2N-bit MLC density does not necessitate additional circuitry but rather involves configuring specific values within the control logic.

It's noteworthy that while N-bit MLC PCMs for N = 3, 4, ... may hold promise for future applications, they might encounter reliability issues such as reduced wear-out endurance and heightened soft error rates due to resistance drift. Consequently, contemplating N-bit MLC PCMs (N > 4) for future product integration proves challenging.

# 6 Conclusion

To improve bit flip reduction and write energy in Phase Change Memory (PCM), we proposed an efficient coding-based data storage, named WIRE, in this paper. WIRE relies on using frequent value locality phenomena. To identify relative occurrence frequency of data patterns, we developed a frequent value finder logic and detected most frequent value locality stack (MFVs). We then design a new coding mechanism that assign specific codes to the identified MFVs with the hamming distance of one. The proposed code guarantees fast access and reduces the bit flip problem. We also optimized our coding mechanism by exploiting partitioning and rotation of the assigned codes with the goal of finding more similarity between the new data to be written and the stored data. Compared to the other bit flip reduction methods, our experimental results showed considerable improvement in IntraV (up to 25%), write energy (up to 32%), and PCM lifetime (up to 20%).

As a future work, we plan to train machine learning algorithms to predict and mitigate bit flips in PCM-based memory systems. By analyzing historical data and identifying patterns associated with bit flip occurrences, machine learning models can proactively adjust write parameters to reduce the likelihood of bit flips during write operations. This can help us predict optimal write parameters for minimizing energy consumption while ensuring data integrity and reliability.